\documentclass[ aps,%
floatfix,%
final,%
notitlepage,%
oneside,%
onecolumn,%
nobibnotes,%
nofootinbib,%
superscriptaddress,%
showpacs,%
]%
{revtex4-1}

\usepackage{graphicx}
\usepackage{amsmath}
\usepackage{hyperref}
\usepackage{booktabs}

\newcommand{\Q}{\mathcal{Q}}

\newcommand{\M}{\mathcal{M}}
\newcommand{\JP}{{J/\psi}}
\newcommand{\fL}[1]{f_{#1}}
\newcommand{\phiL}[1]{\phi_{#1}}

\newcommand{\fN}[1]{\tilde{f}_{#1}}
\newcommand{\GeV}{\,\mathrm{GeV}}

\begin{document}

\title{
Leading order NRQCD and Light-Cone Analysis of Exclusive Charmonia Production in Radiative $Z$-boson Decays 
%Leading Twist Contributions to Charmonium Production in Radiative $Z$  boson Decays
}
\author{A. V. Luchinsky}
\email{alexey.luchinsky@ihep.ru}
\affiliation{``Institute for High Energy Physics" NRC ``Kurchatov Institute'', 142281, Protvino, Russia}
\begin{abstract}
The presented paper is devoted to theoretical analysis of charmonium mesons $\eta_c$, $\JP$, $\chi_{c0,1,2}$, and $h_c$ production in exclusive radiative $Z$ boson decays in the frameworks of Nonrelativistic Chromodynamic and amplitude expansion on the light-cone. In the framework of light-cone expansion amplitudes and widths of all processes are described with one universal relation. It is interesting to note that from this relation it follows automatically which component of $Z$ boson (either vector or axial) gives contribution to the width of the particular decay. It is shown that the effect of internal quark motion almost doubles theoretical predictions for  production of $S$-wave charmonia and $\chi_{c1}$ meson.
\end{abstract}
\pacs{
14.40.Pq, %Heavy quarkonia
12.39.St,	%Factorization
12.39.Jh,	%Nonrelativistic quark model
3.38.Dg	%Decays of Z bosons
}
\maketitle

\section{Introduction}
\label{sec:introduction}

Interplay of long-distance and short-distance effects in Quantum Chromodynamic (QCD) is a long-standing puzzle in the physics of elementary particles. Because of the asymptotic freedom at short distances strong coupling constant $\alpha_s$ is small, so the perturbation theory can be used to describe this region. At the long distances, on the other hand, $\alpha_s\sim 1$ and some nonpeturbative methods should be used.

In the field of heavy quarkonia production and decays one of such methods is the Nonrelativistic Quantum Chromodynamic (NRQCD) \cite{Bodwin:1994jh}, that exploits the fact, that heavy quark's $m_Q$ is large in comparison with the typical scale of the strong interaction $\Lambda_\mathrm{QCD}$. As a result quark's internal velocity inside the meson $v$ should be expected to be small and the amplitude of the process can be written as a series over this parameter. At the leading order (LO) of such expansion internal motion of the quarks is neglected completely and the coefficient, that describe the long-distance part of the interaction can be obtained, for example, using different potential models or comparison with the experiment. It should be noted, however, that in the case of charmonia production this expansion parameter is actually not small ($v^\sim\alpha_c(m_c)\sim 0.3$) so there could be significant errors at the leading order. As an example one can name double charmonia production in $e^+e^-$ annihilation at $B$-factories Belle and BaBar \cite{Abe:2004ww,Aubert:2005tj}, when LO  NRQCD predictions turned out to be about an order of magnitude smaller that the experimental results.

An alternative method for studying charmonia production at high energy interactions is the formalism of amplitude expansion on the light cone (LC) \cite{Chernyak:1983ej}. In this method the internal motion of the quark is described by nonpertubative distribution amplitudes and the small expansion parameter is a chirality factor $m_q/E$, where $m_q$ and $E$ are mass of the produced quark and typical energy of the reaction respectively. Such approach, obviously, is best suitable for description of light meson ($\pi$, $\rho$, etc.) production, but it can be used also in charmonium meson production at high energies. For example, mentioned above disagreement in double charmonia production at $B$-factories was explained in the LC framework \cite{Bondar:2004sv,Braguta:2005kr}. The other example is double charmonium production in exclusive bottomonia decays \cite{Braguta:2009df, Braguta:2009xu, Chen:2012ih}. 

The other interesting type of such processes is exclusive charmonium production in radiative $Z$ boson decays. Up to now this reaction was studied in the literature (see, for example, \cite{Guberina:1980dc}), but usually NRQCD model was used and only leading order results are available. In our paper we will study it in the framework of LC expansion and check to influence of internal quark motion on the branching fractions of the considered decays.

The rest of the paper is organized as follows. In the next section analytical results obtained in LC formalism are presented. Section \ref{sec:nrqcd} is devoted to NRQCD analysis of the considered processes. Numerical predictions for the branching fractions are shown in section \ref{sec:numerical-results} and the last section is reserved for conclusion.

\section{Light Cone Expansion}
\label{sec:light-cone-expansion}

% Общие слова про LC
In this section we will analyze exclusive charmonium $\Q$ production in radiative $Z$ boson decays in the framework of light-cone expansion formalism \cite{Chernyak:1983ej}. In this approach the amplitude of the considered process is written as a series over a small light quark mass $m_q$. It is clear, that this model is most suitable for description of light meson production in high-energy experiments. Recently, however, an attempts were made to use it also for heavy quarkonia production at B-factories Belle and BaBar at $\sqrt{s_{ee}}=10.6$ GeV (see \cite{Abe:2004ww, Bondar:2004sv, Aubert:2005tj, Braguta:2005kr}) and agreement between obtained theoretical predictions and experimental data shows us that light-cone formalism can be used also to describe reactions with heavy quarkonium mesons. Since in $Z$-boson decays the characteristic energy $M_Z\gg\sqrt{s_{ee}}$, one can safely use this model for analysis of charmonium production in these reactions. In the following we will restrict ourselves to the leading order in the expansion over the chirality parameter $r=m_c/M_Z\sim0.02$ (so called leading twist approximation). 
% Формула светки 
Thus, the amplitude of the considered process can be written as a convolution
\begin{align}
  \label{eq:LC_main}
  \M(Z\to\Q\gamma) &\sim f^\Q \int_{-1}^1 d\xi \phi^\Q(\xi) H(\xi),
\end{align}
where $\xi=x-\bar{x}$ is the difference between momentum fractions carried by quark and antiquark respectively,  $H(\xi)$ is a short-distance part of the amplitude, that can be calculated perturbativelly, while leptonic constant $f^\Q$ and distribution amplitude (DA) $\phi^\Q(\xi)$ describe long-distance effects and should be studied using non-perturbative methods such as QCD sum rules.

 According to helicity suppression rules of LC \cite{Chernyak:1983ej} at leading-twist total hadronic helicity in the process should be conserved. Since there is only one hadron in the studied reaction, this rule implies that the helicity of this meson should be $\lambda_\Q=0$, i.e. in leading twist approximation only longitudinally polarized mesons should be produced. The violation of this rule leads to suppression of the amplitude by  small factor of the order $r^{|\lambda_\Q|}$.
% Определение DAs и констант  
The distribution amplitudes that enter in \eqref{eq:LC_main} are defined as
\begin{align}
  \langle\Q_L(p)
  \left|\bar{c}^i_\alpha(z)[z,-z]c^j_\beta(-z)\right|0\rangle
  &=
    \left(\hat{p}\right)_{\alpha\beta}\frac{\fL{\Q}}{4}
    \frac{\delta^{ij}}{3}
    \int\limits_{-1}^1 \phiL{\Q}(\xi)d\xi
\end{align}
for $\Q=\JP$, $\chi_{0,2}$ and
\begin{align}
  \label{eq:DA_odd}
  \langle\Q_L(p)
  \left|\bar{c}^i_\alpha(z)[z,-z]c^j_\beta(-z)\right|0\rangle
  &=
    \left(\hat{p}\gamma_5\right)_{\alpha\beta}\frac{\fL{\Q}}{4}
    \frac{\delta^{ij}}{3}
    \int\limits_{-1}^1 \phiL{\Q}(\xi)d\xi
\end{align}
for $\Q=\eta_c$, $\chi_{c1}$, and $h_c$. In the above expressions $\alpha,\beta$ and $i,j$ are spinor and colour indices of quark and antiquark respectively. It is convenient to define for charmonium mesons so called naturality quantum number
\begin{align}
  \label{eq:narurality}
  \sigma &=P(-1)^J,
\end{align}
where $P$ and $J$ is the space parity and total spin of the particle respectively. One can see, that in the case of negative naturality $\gamma_5$ matrix enters into the definition of DA \eqref{eq:DA_odd}, while in the case $\sigma=1$ no such matrix is present.

 The normalization condition for the distribution amplitudes is
\begin{align}
  \label{eq:norm_xi_even}
  \int_{-1}^1\phiL{\Q}(\xi)d\xi &=1,\qquad   \int_{-1}^1 \xi\phiL{\Q}(\xi)d\xi=1 
\end{align}
for $\xi$-even  ($\Q=\eta_c,\JP, \chi_{c1}$)  and $\xi$-odd ($\Q=\chi_{c0,2}, h_c$) states.
It is easy to check that the parity under $\xi$ inversion is connected with the other quantum numbers of the particle by the relation
\begin{align}
  \label{eq:Pxi}
  P_\xi &=-C\sigma,
\end{align}
where $C$ is the charge parity of the particle.
Explicit parameterization of the distribution amplitudes can be found, for example in papers \cite{Braguta:2006wr, Braguta:2007fh, Braguta:2008qe, Ding:2015rkn, Hwang:2009cu} and we will discuss them in detail in section \ref{sec:numerical-results}.
 For further usage we collect all mentioned above quantum numbers in Table \ref{tab:quantum_numbers}.
% Нормировка, delta-приближение
If internal motion of quarks in meson is neglected (in the following we will refer to this limit as $\delta$-approximation) the distribution amplitudes $\phi_L^\Q(\xi)$ take the form
\begin{align}
  \label{eq:delta_approximation}
  \phiL{\eta_c,\JP,\chi_{c1}}(\xi) &= \delta(\xi),\quad \phiL{\chi_{c0,2},h_c}(\xi) = -\xi\delta'(\xi)
\end{align}
 of $\xi$-even and $\xi$-odd particles respectively. 

\begin{table}
  \centering
  \caption{Quantum numbers of charmonium mesons}
%\begin{tabular}{|c|c|c|c|c|c|}
\begin{tabular}{@{}lllllr@{}}
\hline
 & $J^{PC}$ & $L$ & $S$ & $\sigma$ & $P_{\xi}$\\
\hline
$\eta$ & $0^{-+}$ & 0 & 0 & - & +\tabularnewline
$V$ & $1^{--}$ & 0 & 1 & + & +\tabularnewline
$\chi_{0}$ & $0^{++}$ & 1 & 1 & + & -\tabularnewline
$\chi_{1}$ & $1^{++}$ & 1 & 1 & - & +\tabularnewline
$\chi_{2}$ & $2^{++}$ & 1 & 1 & + & -\tabularnewline
$h$ & $1^{+-}$ & 1 & 0 & - & -\tabularnewline
\hline
\end{tabular}
  \label{tab:quantum_numbers}
\end{table}

\begin{figure}
  \centering
  \includegraphics[width=0.9\textwidth]{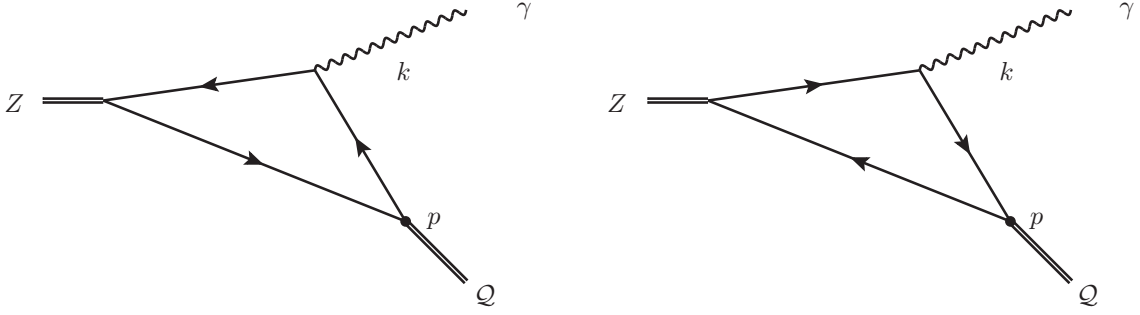}
  \caption{Feynman diagrams for $Z\to\Q\gamma$ process}
  \label{fig:diags}
\end{figure}

As it was noted above, the hard part of the matrix element can be calculated using perturbative QCD. At the leading order this element is described by shown in Fig.~\ref{fig:diags} diagrams. 
% Вершина Z
It is convenient to define $Z\to c\bar{c}$ vertex as
\begin{align}
  \label{eq:Zvertex}
  c_V \gamma_\mu + c_A \gamma_\mu\gamma_\alpha,
\end{align}
where
\begin{align}
  \label{eq:cVcA}
  c_V &=\frac{e}{\sin2\theta_W}\left(\frac{1}{2}-2e_c\sin^2\theta_W\right), \quad 
  c_A=\frac{e}{2\sin2\theta_W}
\end{align}
are the coupling constants of vector and axial components of $Z$-boson and $\theta_W$ is the Weinberg angle. With the above corresponding to these diagrams matrix element for $\sigma$-even meson production takes the form
\begin{align}
  \label{eq:LCmatr}
  \M(Z\to\gamma\Q) &=\frac{4\fL{\Q}}{M_Z^2}e e_c
                     \left\{
                     c_V J_{\mu\nu} \sqrt{I^{(-)}_\Q} + c_A E_{\mu\nu} \sqrt{I^{(+)}_\Q}
                     \right\}\epsilon_Z^\mu\epsilon_\gamma^\nu,
\end{align}
where tensors
\begin{align}
  \label{eq:JE}
  J_{\mu\nu} &= p_\mu k_\nu + p_\nu k_\mu -\frac{M_z^2}{2}g_{\mu\nu},
\quad
  E_{\mu\nu} = e_{\mu\nu\alpha\beta} p^\alpha k^\beta
\end{align}
were introduced and the coefficients $I_\Q^{(\pm)}$ are equal to
\begin{align}
  \label{eq:II}
  \sqrt{I^{(\pm)}_\Q} &= 
                 \frac{1}{2}\int_{-1}^1d\xi \phiL{Q}(\xi)
                 \left[
                 \frac{1}{1-\xi} \pm \frac{1}{1+\xi}
                 \right].
\end{align}
 There is no interference between two terms in eq.~\eqref{eq:LCmatr}, so the width of $\sigma$-even charmonium $\Q$ production reaction is equal to
\begin{align}
  \label{eq:LCgamma}
  \Gamma(Z\to\gamma\Q) &= \frac{2\alpha e_c^2}{3} \frac{\fL{\Q}^2}{M_Z}
                         \left[
                         c_V^2 I^{(-)}_\Q + c_A^2 I^{(+)}_\Q
                         \right].
\end{align}
In the case of $\sigma$-odd meson production one should interchange
$c_{V,A}$ constants in expressions \eqref{eq:LCmatr},
\eqref{eq:LCgamma}. At the leading order these relation agrees with the results presented in \cite{Jia:2008ep,Wang:2013ywc,Grossmann:2015lea, Alte:2017ycm, Alte:2017ycm}.

 It is evident, that coefficients $I^{(-)}_\Q$ and $I^{(+)}_\Q$ are zero for $\xi$-even and $\xi$-odd states respectively. As a result, only one term in equations \eqref{eq:LCmatr}, \eqref{eq:LCgamma} actually gives contribution. For example, in the case of $\JP$-meson production the matrix element of the process is proportional to $c_A$ constant, i.e. only axial component of $Z$ boson is contributing. One should expect this result since, according to charge parity conservation, only axial component of $Z$ boson should give contribution to $Z\to J/\psi(1^{--}) \gamma({1^{--}})$ reaction. It is easy to check, that the same holds also for all other final states: $\sigma$ and $\xi$-parities are consistent with charge-parity conservation.

In $\delta$-approximation \eqref{eq:delta_approximation} non-zero coefficient in \eqref{eq:II} are actually equal to 1, so the width of the considered process takes a simple form
\begin{align}
  \label{eq:Gamma_delta}
  \Gamma_\delta(Z\to\Q\gamma) &= \frac{2\alpha e_c^2}{3}c_{V,A}^2\frac{\fL{\Q}^2}{M_Z},
\end{align}
where, as it was discussed above, the constant $c_{V,A}$ is selected according to charge parity of the final charmonium meson.

\section{NRQCD}
\label{sec:nrqcd}

% Общие слова про NRQCD  
 An alternative way to study the heavy quarkonia production in radiative $Z$ boson decays is Nonrelativistic Quantum Chromodynamic (NRQCD) formalism \cite{Bodwin:1994jh}. This model exploits the fact that mass of the heavy quark $m_Q$ is large in comparison with typical QCD scale $\Lambda_\mathrm{QCD}$. As a result the quark velocity the meson $v^2\sim\alpha_s(m_Q)$ is expected to be small and the amplitude of the process is expanded in a series over this parameter. Long-distance part of the amplitude is described by non-perturbative NRQCD matrix elements, that are usually determined from solution of potential models of an analysis of existing experimental data. It should be noted, that in NRQCD framework fock columns of charmonia mesons contain not only color-singlet components $(c\bar{c})_{1_c}$, but also color-octet terms. In the latter case overall color neutrality is enforced by the presence of additional gluons, e.g. $(c\bar{c})_{8_c}g$. In the case of $P$-wave charmonia $\chi_{c}$, $h_c$ color-octet components are expected to give the same order contributions as color-singlet ones. Note, however, that in considered in our article case final charmonium is the only strong-interacting particle, so there is no way for additional gluon to hadronize. For this reason in the following we will take into account only contributions of color-singlet components.

 Using the projection technique presented in \cite{Braaten:2002fi} it is easy to obtain the following expression:
 \begin{align}
   \label{eq:NRQCDmatr}
   \left|\M\left[ Z(\lambda_Z)\to\Q(\lambda_Q)\gamma(\lambda_\gamma\right)]\right|^2 &=
           16\pi\alpha e_c^2\fN{Q}^2 C^\Q_{\lambda_\gamma\lambda_\Q},
 \end{align}
where $\lambda_{Z,\gamma,\Q}$ are the helicities of the corresponding particles (note, that due to orbital momentum conservation $\lambda_Z=\lambda_\Q-\lambda_\gamma$). Analogous to LC leptonic constants $\fL{\Q}$ NRQCD constants $\fN{\Q}$ are defined as
\begin{align}
  \label{eq:f_NRQCD_S}
 \fN{\eta_c} &= \fN{\JP} = \sqrt{\frac{\langle O_1\rangle_{\JP}}{m_c}}
\end{align}
for $S$-wave mesons and
\begin{align}
  \label{eq:f_NRQCD_P}
 \fN{h_c} &= \sqrt{3} \fN{\chi_{c0}} = \frac{1}{\sqrt{2}}\fN{\chi_{c1}} = \sqrt{\frac{2}{3}} \fN{\chi_{c2}} = \sqrt{\frac{\langle O_1\rangle_{h_C}}{m_c^3}},
\end{align}
for $P$-waxe states. Definitions of the matrix elements $\langle O_1\rangle_{\JP,h_c}$ can be found  in \cite{Braaten:2002fi}. As for $C_{\lambda_\gamma\lambda_\Q}^\Q$ coefficients in eq.~\eqref{eq:NRQCDmatr}, from space symmetry it follows that $C^\Q_{-\lambda_\gamma,-\lambda_\Q}=C^\Q_{\lambda_\gamma,\lambda_\Q}$ and nonzero coefficients are equal to
\begin{align}
C^{\eta_c}_{10}&= 1
\end{align} 
for $\eta_c$ meson,
\begin{align}
C^{J/\psi}_{10}&= 1,\quad C^{J/\psi}_{11} = 4 r^2
\end{align}
for $J/\psi$,
\begin{align}
C^{\chi_{c0}}_{10}&= \frac{\left(1-12 r^2\right)^2}{\left(1-4 r^2\right)^2}
\end{align}
for $\chi_{c0}$,
\begin{align}
C^{\chi_{c1}}_{10}& = \frac{1}{\left(1-4 r^2\right)^2},\quad
C^{\chi_{c1}}_{11} = \frac{4 r^2}{\left(1-4 r^2\right)^2}
\end{align}
for $\chi_{c1}$,
\begin{align}
C^{\chi_{c2}}_{10}&=  \frac{1}{\left(1-4 r^2\right)^2}, \quad
C^{\chi_{c2}}_{11}= \frac{12 r^2}{\left(1-4 r^2\right)^2},\quad
C^{\chi_{c2}}_{12}= \frac{96 r^4}{\left(1-4 r^2\right)^2}
\end{align}
for $\chi_{c2}$ and
\begin{align}
C^{h_c}_{10}&=  1,\quad
C^{h_c}_{11}=  4 r^2.
\end{align}
in the case of $h_c$ production. 
In the above expression the chirality factor
\begin{align}
  r=\frac{m_c}{M_Z}
\end{align}
was introduced. It is clear that helicity suppression rule $C^\Q_{\lambda_\gamma\lambda_\Q}\sim r^{2|\lambda_Q|}$ holds and main contributions come from  production of  longitudinally polarized charmonium meson.

% Выражения для ширин  
 Summed over polarizations widths of the considered processes are equal to
\begin{align}
  \label{eq:NRQCDwidths}
\Gamma(Z\to\gamma \eta_c) &=\frac{2 \alpha  c_V^2 e_c^2 \fN{\eta_c}^2 }{3 M_Z}\left(1-4 r^2\right),\\
\Gamma(Z\to\gamma J/\psi) &=\frac{2 \alpha  c_A^2 e_c^2 \fN{J/\psi}^2 }{3 M_Z}\left(1-16 r^4\right),\\
\Gamma(Z\to\gamma \chi_{c0}) &=\frac{2 \alpha  c_V^2 e_c^2 \fN{\chi_{c0}}^2}{3 M_Z}\frac{\left(1-12 r^2\right)^2}{ 1-4 r^2},\\
\Gamma(Z\to\gamma \chi_{c1}) &=\frac{2 \alpha  c_V^2 e_c^2 \fN{\chi_{c1}}^2}{3 M_Z }\frac{1+4 r^2}{1-4 r^2},\\
\Gamma(Z\to\gamma \chi_{c2}) &=\frac{2 \alpha  c_V^2 e_c^2 \fN{\chi_{c2}}^2}{3 M_Z}\frac{1+12 r^2+96 r^4}{1-4 r^2}\\
\Gamma(Z\to\gamma h_c) &=\frac{2 \alpha  c_A^2 e_c^2 \fN{h_c}^2}{3 M_Z}\left(1-16 r^4\right).
\end{align}
% В ведущем твисте совпадает с LC  
 It is clear, that in the limit $r\to0$ these cross sections coincide with $\delta$-approximation of LC result \eqref{eq:LCgamma}, where meson distribution amplitudes are chosen in the form \eqref{eq:delta_approximation}. Presented above expressions agree with the results given in \cite{Guberina:1980dc}.

\section{Numerical Results}
\label{sec:numerical-results}

In this section we present numerical predictions for the branching fractions of the considered processes in the framework of NRQCD and LC models. The mass and total width of $Z$ boson is known experimentally \cite{Olive:2016xmw}:
\begin{align}
  M_Z &= 92\GeV,\quad \Gamma_Z = 2.5\GeV.
\end{align}
In our paper we will use the following expression for the strong coupling constant
\begin{align}
  \label{eq:alphas}
  \alpha_s(\mu^2) &= \frac{4\pi}{b_0\ln(\mu^2/\Lambda_\mathrm{QCD}^20}, \quad b_0=11-\frac{2}{3}n_f,
\end{align}
where $\Lambda_\mathrm{QCD}\approx 0.2\,\GeV$ and $n_f=5$ is the number of active flavors. At the scale $\mu^2=M_Z^2$ it corresponds to $\alpha_s(M_Z^2)\approx 0.13$.

Let us consider first theoretical predictions in the framework of NRQCD model. At the leading order of perturbation theory the matrix elements $\langle O_1\rangle_{\JP,\chi}$ can be obtained from the decay width of the corresponding mesons:
\begin{align}
  \Gamma(\JP\to\mu\mu) &= \frac{2\pi\alpha e_c^3}{3}\frac{\langle O_1\rangle_{\JP}}{m_c^2},
\quad
  \Gamma(\chi_{c2}\to2\gamma) = \frac{8\pi e_c^4\alpha^2}{5}\frac{\langle O_1\rangle_\chi}{m_c^4},
\end{align}
with the $c$-quark mass $m_c=M_\JP/2\approx 1.5$ GeV. Using experimental results for the above widths it is easy to obtain the values of NRQCD matrix elements \cite{Braaten:2002fi}:
\begin{align}
  \label{eq:ONRQCD_num}
  \langle O_1\rangle_\JP &=0.222\,\GeV^2, \quad \langle O_1\rangle_\chi = 0.033\,\GeV^5.
\end{align}
These values correspond to NRQCD leptonic constants \eqref{eq:f_NRQCD_S}, \eqref{eq:f_NRQCD_P} are equal to
\begin{align}
  \label{eq:fNRQCD_num}
 \fN{\eta_c} &= \fN{\JP}=0.38,\GeV,\\
  \fN{\chi_{c0}} &= 0.057\,\GeV, \fN{\chi_{c1}}=0.14,\GeV, \\
  \fN{\chi_{c2}} &=0.081\,\GeV, \fN{h_c}=0.099\,\GeV.
\end{align}
Obtained with these parameters NRQCD predictions are presented in the second column of the table \ref{tab:tab}.

In order to calculate branching fractions under LC approach we need to know numerical values of the leptonic constants and distribution amplitudes. This question was studied thoroughly in the literature, in the following we will use results presented in papers \cite{Braguta:2006wr, Braguta:2007fh, Braguta:2008qe}. In particular, calculated on the scale $\mu=m_c$ leptonic constants of charmonia mesons are taken to be equal to
\begin{align}
  \label{eq:fLC_num}
  \fL{\eta_c}(m_c) &=(0.35\pm0.02)\GeV, \quad \fL{\JP}(m_c)=(0.41\pm0.02)\GeV,\quad
  \fL{\chi_{c0}}(m_c) = (0.11\pm0.02)\GeV,\\
 \fL{\chi_{c1}}(m_c) &= (0.27\pm0.05)\GeV,
  \fL{\chi_{c2}}(m_c)  = (0.16\pm0.03)\GeV, \quad \fL{h_c}(m_c) = (0.19\pm0.03)\GeV.
\end{align}
Note, the value of $c$ quark in LC model differs from phenomenological choice $M_{\JP}/2$ and is equal to $m_c=1.2\GeV$.

In the case of $\delta$-approximation distribution amplitudes are
given by the relations \eqref{eq:delta_approximation} and the widths
of the decays are given by relation \eqref{eq:Gamma_delta}. The
corresponding results are presented in the third column of table
\ref{tab:tab}. The errors in this column stems from uncertainties of
the leptonic constants. One can easily see, that LC predictions in
$\delta$-approximation differ a little bit from NRQCD ones. Because of
the smallest of the chirality factor $r=m_c/M_Z\sim 0.02$ charm quark mass correction are small and this difference is explained by the difference between leptonic constants \eqref{eq:fNRQCD_num} and  \eqref{eq:fLC_num}.

In order to calculate the branching fractions in full LC framework some parameterization for the functions $\phi_L^\Q(\xi)$ is required. Before we proceed to this point it is worth noting one important thing. The distribution amplitudes in eq.~\eqref{eq:LC_main} actually depend on the scale $\mu$, which is usually taken of the order of typical energy of the reaction. Presented in the literature results are given at the scale $\mu_0=m_c$ so we need to track the evolution of these functions to $\mu=M_Z$. In order do this it is convenient to write the function as a series over Gegenbauer polynomials
\begin{align}
  \label{eq:Gegenbauer}
  \phiL{\Q}(\xi,\mu_0) &= \sum_{n=n_0}^\infty a_n^\Q(\mu_0) C_n^{3/2}(\xi). 
\end{align}
It is clear, that for $\xi$-even and $\xi$-odd distribution amplitudes only even and odd coefficients are different from zero, so $n_0=0$ and $1$ in these two cases. According to \cite{Gribov:1972rt, Lipatov:1974qm, Altarelli:1977zs, Braun:2003rp} evolution results to the change of coefficients
\begin{align}
  a_n^\Q(\mu) &= L^{-\frac{\gamma_n}{b_0}} a_n^\Q(\mu_0),
\end{align}
where anomalous dimensions $\gamma_n$ for longitudinal current are equal to
\begin{align}
  \gamma_n = \frac{4}{3}\left(1-\frac{2}{(n+1)(n+2)}+\sum_{j=2}^{n+1}\frac{1}{j}\right)
\end{align}
and
\begin{align}
 L &= \frac{\alpha_s(\mu)}{\alpha_s(\mu_0)}.
\end{align}
As a result of this evolution the width of the distribution amplitude grows, but the normalization conditions \eqref{eq:norm_xi_even} are violated. If one wants to keep this normalization, it is possible to factor out the scale-dependent factor from the first term in the expansion \eqref{eq:Gegenbauer} and move it to the corresponding leptonic constant. As a result of this procedure the normalization of distribution amplitudes restores, but the evolution of the corresponding leptonic constant appears
\begin{align}
\phiL{\Q}(\xi,\mu) &= \sum_{n=n_0}^\infty 
                L^{-\frac{\gamma_n - \gamma_{n_0}}{b_0}}
a_n^\Q(\mu_0) C_n^{3/2}(\xi),
\quad
\fL{\Q}(\mu) = L^{-\frac{\gamma_{n_0}}{b_0}} \fL{\Q}(\mu_0).
\end{align}
 Since the first anomalous dimension $\gamma_0=0$, in the case of $\xi$-even states $\eta_c$, $J/\psi$, and $\chi_{c1}$ this evolution does change the constants. In the case of $\xi$-odd states $\chi_{c0,2}$, $h_c$, on the other hand with the increase of the scale leptonic constant decreases.

\begin{figure}
  \centering
  \includegraphics[width=0.9\textwidth]{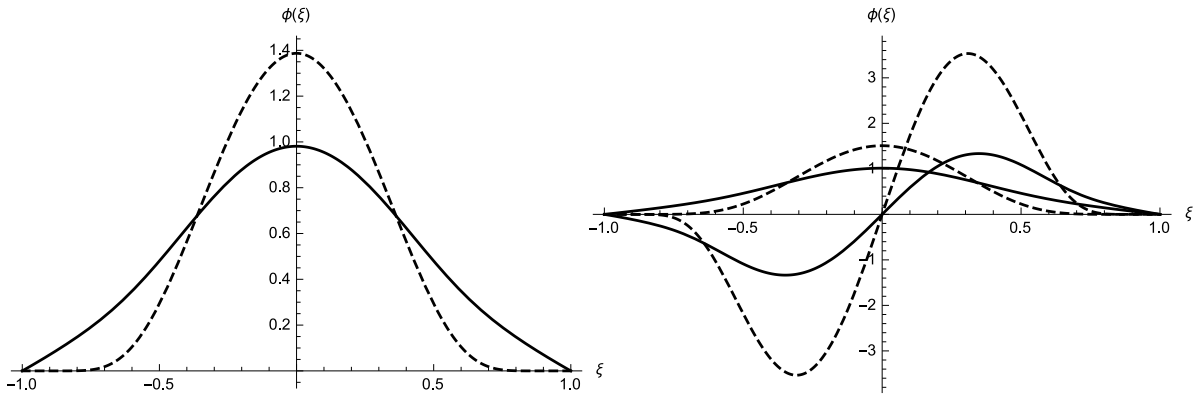}
  \caption{Distribution amplitudes of the $S$-wave (left figure) and $P$-wave (right figure) charmonium mesons. Dashed and solid lines correspond to $\mu^2=m_c^2$ and $\mu^2=M_Z^2$ scales respectively. In the case of $P$-wave states $\xi$-even curve correspond to $\chi_{c1}$ meson, while $\xi$-odd one to $\chi_{c0,2}$, $h_c$}
  \label{fig:wf}
\end{figure}

According to papers \cite{Braguta:2006wr, Braguta:2007fh} the distribution amplitude of $S$-wave charmonia can be written in the form
\begin{align}
  \phiL{\JP,\eta_c}(\xi,\mu_0) &= c(\beta_S) (1-\xi^2) \exp\left(-\frac{\beta_S}{1-\xi^2}\right),
\end{align}
where $c(\beta_S)$ is a normalization constant \eqref{eq:norm_xi_even} and
\begin{align}
  \beta_S &=3.8\pm0.7.
\end{align}
In the case of $P$-wave states there are two distribution functions \cite{Braguta:2008qe}:
\begin{align}
  \label{eq:6}
  \phiL{\chi_{c0,2},h_c}(\xi,\mu_0) &= c_1(\beta_P) \xi(1-\xi^2) \exp\left(-\frac{\beta_P}{1-\xi^2}\right),
\end{align}
for $\xi$-odd states and
\begin{align}
  \phiL{\chi_{c1}}(\xi,\mu_0) &= -c_2(\beta_P)\int_{-1}^\xi \phiL{h_c}(\xi,\mu_0)
\end{align}
for $\xi$-even. In these expressions $c_{1,2}(\beta_P)$ is the normalization constant \eqref{eq:norm_xi_even} and
\begin{align}
  \beta_P &=3.4^{+1.5}_{-0.9}.
\end{align}
In figure \ref{fig:wf} we show these distribution amplitudes at different scales. One can see that with the increase of the scale the effective widths of the DAs increase.
\begin{table}
  \centering
%\begin{tabular}{|c||c|c|c|c|}
\begin{tabular}{@{}llllr@{}}
\hline
 $\Q $ &  $Br_\mathrm{NRQCD},\, 10^{-8}$ &  $Br_\delta,\, 10^{-8}$ &  $Br_\mathrm{LC},\, 10^{-8}$  & $Br_{LC}/Br_\delta$ \\ 
\hline 
$\eta_c$ & $0.66$ &  $0.55\pm0.08_\mathrm{f}$ &  $0.94\pm0.1_\mathrm{f} \pm 0.01_\mathrm{wf}$ &  $1.7\pm0.02_\mathrm{wf}$ \\ 
$J/\psi$ & $4.5$ &  $5.1\pm0.5_\mathrm{f}$ &  $8.8\pm0.9_\mathrm{f} \pm 0.09_\mathrm{wf}$ &  $1.7\pm0.02_\mathrm{wf}$ \\ 
$\chi_{c0}$ & $0.014$ &  $0.055\pm0.02_\mathrm{f}$ &  $0.05\pm0.02_\mathrm{f} \pm 0.003_\mathrm{wf}$ &  $0.91\pm0.05_\mathrm{wf}$ \\ 
$\chi_{c1}$ & $0.087$ &  $0.33\pm0.1_\mathrm{f}$ &  $0.56\pm0.2_\mathrm{f} \pm 0.007_\mathrm{wf}$ &  $1.7\pm0.02_\mathrm{wf}$ \\ 
$\chi_{c2}$ & $0.029$ &  $0.11\pm0.04_\mathrm{f}$ &  $0.1\pm0.04_\mathrm{f} \pm 0.006_\mathrm{wf}$ &  $0.91\pm0.05_\mathrm{wf}$ \\ 
$h_c$ & $0.3$ &  $1.1\pm0.4_\mathrm{f}$ &  $1.\pm0.4_\mathrm{f} \pm 0.06_\mathrm{wf}$ &  $0.91\pm0.05_\mathrm{wf}$ \\ 
\hline 
\end{tabular}
\caption{Branching fractions of radiative decays $Z\to\Q\gamma$}
\label{tab:tab}
\end{table}

The results obtained using presented above distribution amplitudes are shown in the fourth column of table~\ref{tab:tab}. The errors labeled with subscript ``$\mathrm{wf}$'' in this column are caused by uncertainties in DAs parameters $\beta_{S,P}$. In order to show the effect of internal motion of the quarks in mesons in the fifth column of that table we present the ratio of branching fractions obtained in $\delta$-approximation and full LC. One can see, that in the case of $\xi$-even  charmonia this effect almost doubles the branching fraction, while for $\xi$-odd states the branching fractions are slightly decreased. This difference is explained by mentioned above fact that during the evolution leptonic constants $\fL{\Q}$ do not change and decrease in the former and latter cases.  In the case of $J/\psi$-meson production our result is in reasonable agreement with the papers \cite{Grossmann:2015lea, Alte:2017ycm}.

\section{Conclusion}
\label{sec:conclusion}

The presented paper is devoted to theoretical analysis of charmonium meson production in exclusive radiative $Z$ boson decays in the frameworks of Nonrelativistic Chromodynamic and amplitude expansion on the light-cone. It is shown that, in agreement with LC selection rules longitudinally polarized mesons give main contribution to the branching fraction and the effect of internal quark motion increases almost doubles theoretical predictions for some of the processes (i.e. production of $S$-wave charmonia and $\chi_{c1}$ meson). It should also be  noted, that because of nontrivial cancellation of various terms in the  amplitude space parity of the light-cone distribution functions consistent with charge parity of the final meson component of $Z$ boson (vector or axial) gives contributions. 

Resulting branching fractions are still rather small ($\sim 10^{-8}$),
so the considered processes can be considered as rare $Z$ boson
decays. They are, however, comparable with the branching fractions of
some experimentally observed decays (e.g. $B_s\to\mu\mu$
\cite{CMS:2014xfa,Bobeth:2013uxa}), so one could expect that it is
also possible to observe them. Experimental measurements of the
discussed branching fractions could help us to understand better the
physics of charmonium mesons. It is interesting to note, that
according paper work \cite{Achasov:1991iv} in the case of $\JP$ meson
production vector dominance model results in significant increase of
the corresponding branching fraction.

It is worth mentioning that in the recent paper \cite{Ma:2017xno} a new Soft Gluon Factorization method (SGF) was proposed to describe charmonia production processes. In this approach velocity expansion is treated with more accuracy, so this model could be more suitable for calculations in comparison with NRQCD approach. In our future work we plan to study considered in our paper decays in SGF framework.

The author would like to thank A.K. Likhoded for fruitful discussions.

% \bibliographystyle{apsrev4-1}
% \bibliography{Z_QQgamma}

%

\end{document}